# A polynomial-time heuristic for Circuit-SAT

Francesco Capasso

capassofrancesco@yahoo.it

**Abstract.** In this paper is presented an heuristic that, in polynomial time and space in the input dimension, determines if a circuit describes a tautology or a contradiction. If the circuit is neither a tautology nor a contradiction, then the heuristic finds an assignment to the circuit inputs such that the circuit is satisfied.

## 1. Introduction

Circuit - SAT: given a circuit C and $x_1, \ldots, x_n$ inputs find a value 0 or 1 for each input such that circuit's output equals 1.

Circuit-SAT is one of the main NP-complete problems [1], however most solvers proposed are based on the reduction from Circuit-SAT to SAT, which is another NP-complete problem. Many attempts have been made so far in order to efficiently solve SAT, but no satisfying solution has been found yet.

The approach adopted to solve Circuit-SAT follows the ideas already outlined in "*Solving SAT with Bilateral Computing*" [2] and consists in a reverse circuit execution. Beginning from the output, that has a value 1 or 0, we go back through the circuit gates up to its inputs. In so doing we get those input values such that the output equals the value fixed in the beginning.

The present heuristic finds, in polynomial time and space in the input dimension, an assignment, if exists, to the inputs of a circuit C such that the output equals 1.

In what follows is introduced a list of logic gates features (2) and two data structures (3). Subsequently is presented the heuristic (4) and its proof of correctness (5) for some particular circuit classes. Then is introduced a recursive function (6) to update data structures, after that is given the correctness of heuristic for each circuit class (7) and then is calculated its computational complexity (8).

## 2. Properties of logic gates

Given a circuit C we assign a label $x_i$ to each input ($0 \leq i \leq n$), a label $G_j$ to each logic gate and a label $y_j$ to each logic gate output ($0 \leq j \leq m$).

In this way the circuit in figure 1 can be represented by formula 1.

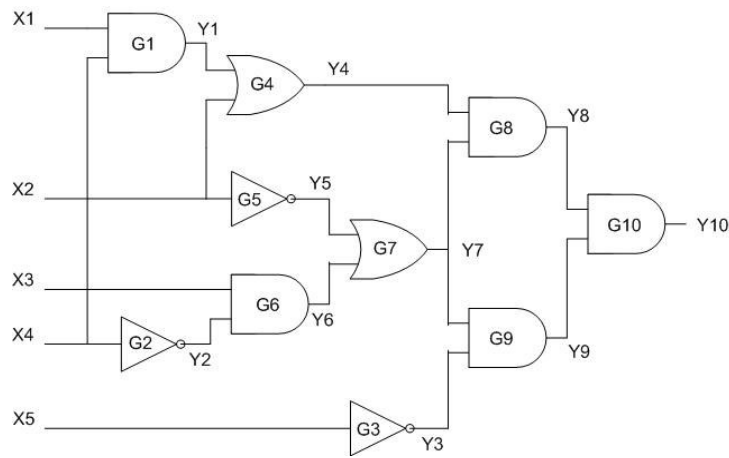

**Figure 1.** Sample circuit.

$$(y_1 = x_1 \wedge x_4) \wedge (y_2 = \neg x_4) \wedge (y_3 = \neg x_5) \wedge (y_4 = y_1 \vee x_2) \wedge (y_5 = \neg x_2) \wedge$$
$$\wedge (y_6 = x_3 \wedge y_2) \wedge (y_7 = y_5 \vee y_6) \wedge (y_8 = y_4 \wedge y_7) \wedge \qquad (1)$$
$$\wedge (y_9 = y_7 \wedge y_3) \wedge (y_{10} = y_8 \wedge y_9)$$

We notice that each $y_j$ is associated to the output of a gate $G_j$ whose truth table is known (table 1).

| AND | | | OR | | | NOT | |
|---|---|---|---|---|---|---|---|
| x | y | out | x | y | out | x | out |
| 0 | 0 | 0 | 0 | 0 | 0 | 0 | 1 |
| 0 | 1 | 0 | 0 | 1 | 1 | 1 | 0 |
| 1 | 0 | 0 | 1 | 0 | 1 | | |
| 1 | 1 | 1 | 1 | 1 | 1 | | |

**Table 1.** Truth table for AND, OR and NOT gates.

**Definition**

With the term *degree of freedom* we define the maximum number of possible combinations of values that can be assigned to a logic gate according to the values given to its inputs and output.

Classification of the degrees of freedom:

- Degree 4:
    - AND and OR gate without any specific value mapped on inputs or output;

- Degree 3:
    - AND gate with output 0; OR gate with output 1;

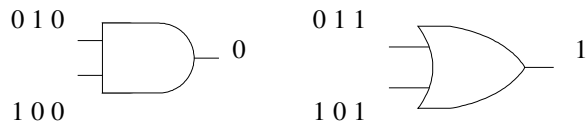

- Degree 2:
    - AND gate with output 0 and an input 0; AND gate with an input 0;

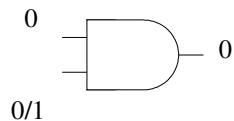

- OR gate with output 1 and an input 1; OR gate with an input 1;

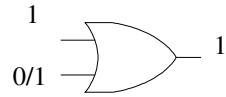

- Degree 1:

    - AND gate with output 0 and input 1; AND gate with output 1;

    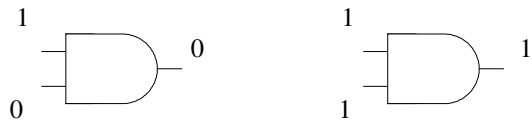

    - OR gate with output 1 and an input 0; OR gate with output 0;

    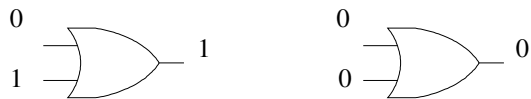

    - NOT gate with output 0 or 1, or input 0 or 1;

    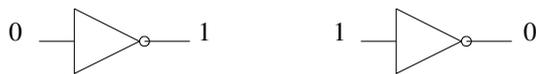

- Degree 0 (contradiction):

    - AND gate with output 1 and an input 0;

    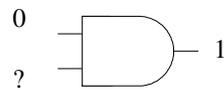

    - OR gate with output 0 and input 1;

    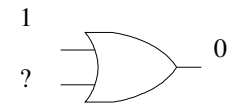

o   NOT gate with output and input both 0 or 1;

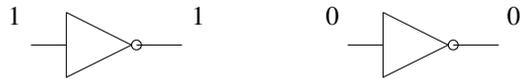

## 3. Data structures

We represent a circuit C using following data structures:

1. a *circuit truth table* (CTT) with $n+m$ columns (labelled by $x_i$ and $y_j$): for each gate we report the corresponding truth table (setting to "null" the fields not assigned to 0 or 1); to each value we assign the label of the gate it belongs.

2. a *degree table* (DT) with $m$ columns (labelled by $G_j$): for each gate we report the index of circuit's truth table rows in which are defined truth value for the gate itself.

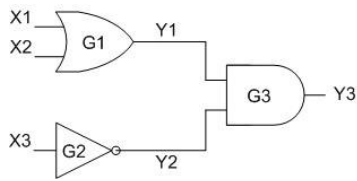

**Figure 2.** Sample circuit.

The circuit featured in figure 2 can be represented by the following formula:

$$(y_1 = x_1 \vee x_2) \wedge (y_2 = \neg x_3) \wedge (y_3 = y_1 \wedge y_2) \tag{2}$$

|    | X1    | X2    | X3    | Y1    | Y2    | Y3    |
|----|-------|-------|-------|-------|-------|-------|
| 1  | $0/G_1$ | $0/G_1$ |       | $0/G_1$ |       |       |
| 2  | $0/G_1$ | $1/G_1$ |       | $1/G_1$ |       |       |
| 3  | $1/G_1$ | $0/G_1$ |       | $1/G_1$ |       |       |
| 4  | $1/G_1$ | $1/G_1$ |       | $1/G_1$ |       |       |
| 5  |       |       | $0/G2$ |       | $1/G_2$ |       |
| 6  |       |       | $1/G_2$ |       | $0/G_2$ |       |
| 7  |       |       |       | $0/G_3$ | $0/G_3$ | $0/G_3$ |
| 8  |       |       |       | $0/G_3$ | $1/G_3$ | $0/G_3$ |
| 9  |       |       |       | $1/G_3$ | $0/G_3$ | $0/G_3$ |
| 10 |       |       |       | $1/G_3$ | $1/G_3$ | $1/G_3$ |

**Table 2.** Circuit truth table of circuit in figure 2.

| $G_1$ | $G_2$ | $G_3$ |
|-----|-----|-----|
| 1   | 5   | 7   |
| 2   | 6   | 8   |
| 3   |     | 9   |
| 4   |     | 10  |

**Table 3.** Degree table of circuit in figure 2.

We observe that in circuit truth table the maximum number of rows we introduce for each gate is 4, consequently the table has in the worst case dimension $(n+m) \cdot (4m)$ that is $O(nm+m^2)$.

Similarly we observe that in degree table the maximum number of rows introduced is 4 and the resulting dimension is therefore $4 \cdot m$ that is $O(m)$.

## 4. Heuristic

Given $0 \leq j \leq m$ and supposing that each $G_j$ is correlated with a $y_j$:

1. assign a value to the output, $y_m=1$;

2. delete from the circuit truth table all rows in which $y_m= 0$; delete from degree table the values related to the index of previous deleted rows in circuit truth table;

3. until unassigned $x_i$ exist:

    3.1. assign missing value to the inputs or output of all gates with degree of freedom 1; update circuit truth table and degree table (as shown in step 2);

    3.2. if there are no more gates with degree of freedom 1, assign the value 0 or 1 to an input of a gate with degree of freedom 2 or 3; then update the circuit truth table and degree table (as shown in step 3.2);

Applying the heuristic to the circuit showed in figure 2 we obtain:

- [step 1]: set $y_3=1$;
- [step 2]:
    - delete from CTT all rows where $y_3 = 0$ (rows 7, 8 and 9);
    - delete from DT, field $G_3$, the values 7, 8 and 9;
- [step 3.1]:
    - degree($G_1$)=4; degree($G_2$)=2; degree($G_3$)=1 $\Longrightarrow$ $y_2$=1 and $y_1$= 1 (row 10);
    - delete from CTT all rows where $y_1 = 0$ (row 1) ed $y_2 = 0$ (row 6);
    - delete from DT, field $G_1$ value 1, and in field $G_2$ value 6;
    - degree($G_1$)=3, degree($G_2$)=1 $\Longrightarrow$ $x_3 = 0$ (row 5);
- [step 3.2]:
    - degree($G_1$)=3;
    - determine a value for one of the inputs in $G_1$, for instance $x_1 = 1$;
    - delete from CTT all rows where $x_1 = 0$ (row 2);
    - delete from DT, field $G_1$, the value 2;
    - degree($G_1$)=2;

- set a value for $x_2$, for instance $x_2 = 0$;
- delete from CTT all rows where $x_2=1$ (row 4);
- delete from DT, field $G_1$, the value 4;

The solution is given by the remaining rows in the circuit truth table (3, 5 e 10). These rows contain the assignment sought for $x_1$, $x_2$ and $x_3$.

## 5. Correctness

The correctness of steps 1 and 2 is obvious. To demonstrate the correctness of steps 3.1 and 3.2 we observe circuit in figure 3. This circuit describes neither a contradiction nor a tautology therefore exists an assignment to its inputs that satisfy the circuit. In the first case (a) applying step 3.1 and then 3.2 we obtain a solution, while in the second case applying step 3.2 before step 3.1 we obtain a contradiction. In other words applying step 3.1 we strongly bind the inputs and output assignment.

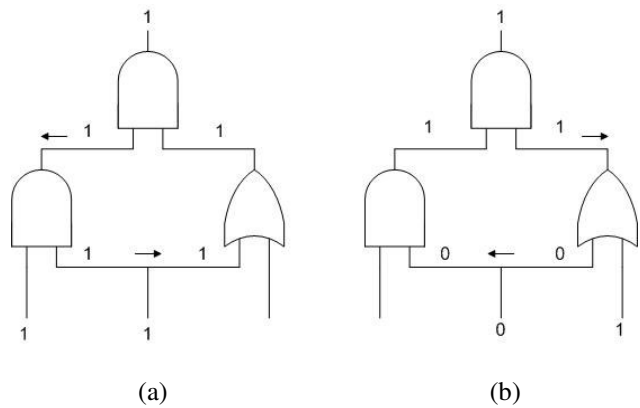

(a)                                                    (b)

**Figure 3.** Solution (a); Contradiction (b).

The execution of step 3.1 brings to 3 different configurations:

- a chain of consistent assignments for every logic gate and input (for instance, that is the case of a circuit with AND gates only);
- a chain of consistent assignments that ends with an inconsistent assignment (that is the case of a circuit which represents a contradiction, figure 4);

- a chain of consistent assignments for some logic gates and inputs with degree of freedom 1, where the other gates have degree of freedom 2 or 3 (figure 5);

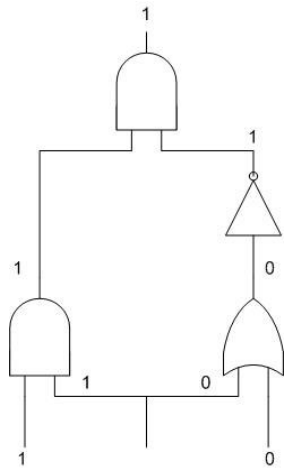

**Figure 4.** Circuit that represent a contradiction.

In the first case the heuristic ends with a solution, whereas in the second case it tells us that a solution does not exist. The third case is more articulated and introduces us to the correctness's proof of step 3.2.

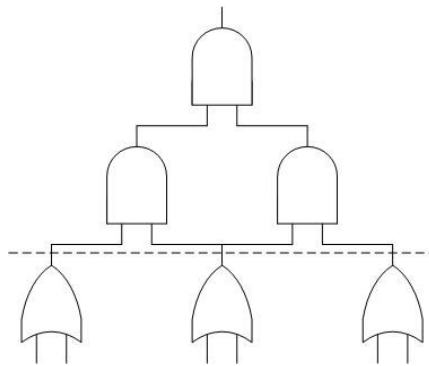

**Figure 5.** Correctness of step 3.2.

Let us suppose that all gates above the dashed line in figure 5 have been resolved in step 3.1, since they are gates with degree of freedom 1, whereas the gates under the same dashed line have degree of freedom 2 or 3.

At this point we could choose any still unsolved gate (under the dashed line) and then set one of its inputs to 0 or 1 (step 3.2).

A random assignment for sure is consistent with the part of the circuit which has already been solved (backward), since any assigned value does not contradict the values already fixed. On the other hand it isn't always consistent with the remaining part of the circuit (forward).

In order to prove that, let us consider the circuit in figure 6.

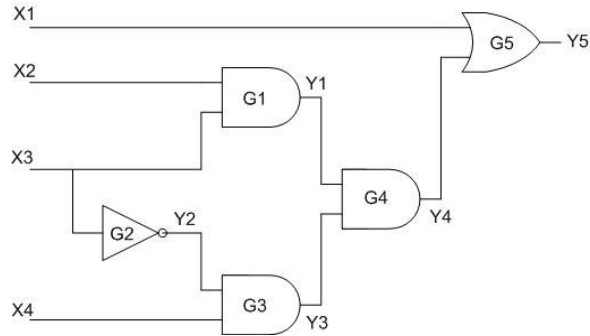

**Figure 6.** Circuit with subcircuit that describes a contradiction.

The circuit in figure 6 can be represented by the following formula:

$$(y_1 = x_2 \wedge x_3) \wedge (y_2 = \neg x_3) \wedge (y_3 = y_2 \wedge x_4) \wedge (y_4 = y_3 \wedge y_1) \wedge (y_5 = y_4 \vee x_1) \qquad (3)$$

|    | X1      | X2      | X3      | X4      | Y1      | Y2      | Y3      | Y4      | Y5      |
|----|---------|---------|---------|---------|---------|---------|---------|---------|---------|
| 1  |         | 0/ $G_1$ | 0/ $G_1$ |         | 0/ $G_1$ |         |         |         |         |
| 2  |         | 0/ $G_1$ | 1/ $G_1$ |         | 0/ $G_1$ |         |         |         |         |
| 3  |         | 1/ $G_1$ | 0/ $G_1$ |         | 0/ $G_1$ |         |         |         |         |
| 4  |         | 1/ $G_1$ | 1/ $G_1$ |         | 1/ $G_1$ |         |         |         |         |
| 5  |         |         | 0/ $G_2$ |         |         | 1/ $G_2$ |         |         |         |
| 6  |         |         | 1/ $G_2$ |         |         | 0/ $G_2$ |         |         |         |
| 7  |         |         |         | 0/ $G_3$ |         | 0/ $G_3$ | 0/ $G_3$ |         |         |
| 8  |         |         |         | 0/ $G_3$ |         | 1/ $G_3$ | 0/ $G_3$ |         |         |
| 9  |         |         |         | 1/ $G_3$ |         | 0/ $G_3$ | 0/ $G_3$ |         |         |
| 10 |         |         |         | 1/ $G_3$ |         | 1/ $G_3$ | 1/ $G_3$ |         |         |
| 11 |         |         |         |         | 0/ $G_4$ |         | 0/ $G_4$ | 0/ $G_4$ |         |
| 12 |         |         |         |         | 0/ $G_4$ |         | 1/ $G_4$ | 0/ $G_4$ |         |
| 13 |         |         |         |         | 1/ $G_4$ |         | 0/ $G_4$ | 0/ $G_4$ |         |
| 14 |         |         |         |         | 1/ $G_4$ |         | 1/ $G_4$ | 1/ $G_4$ |         |
| 15 | 0/ $G_5$ |         |         |         |         |         |         | 0/ $G_5$ | 0/ $G_5$ |
| 16 | 0/ $G_5$ |         |         |         |         |         |         | 1/ $G_5$ | 1/ $G_5$ |
| 17 | 1/ $G_5$ |         |         |         |         |         |         | 0/ $G_5$ | 1/ $G_5$ |
| 18 | 1/ $G_5$ |         |         |         |         |         |         | 1/ $G_5$ | 1/ $G_5$ |

**Table 4.** Circuit truth table of circuit in figure 6.

| G1 | G2 | G3 | G4 | G5 |
|----|----|----|----|----|
| 1  | 5  | 7  | 11 | 15 |
| 2  | 6  | 8  | 12 | 16 |
| 3  |    | 9  | 13 | 17 |
| 4  |    | 10 | 14 | 18 |

**Table 5.** Degree table of circuit in figure 6.

Let us apply the heuristic to the circuit in figure 6:

- [step 1]: $y_5 = 1$;

- [step 2]:

  - delete from the circuit truth table all rows where $y_5 = 0$ (row 15);

  - delete from the degree table, field $G_5$, the value 15;

  - degree($G_1$)=4; degree($G_2$)=4; degree($G_3$)=4; degree($G_4$)=4; degree($G_5$)=3;

- [step 3.2]:

  - …

As the degree of $G_5$ equals 3, a random choice between 0 and 1 is necessary for one of the inputs ($y_4$ or $x_1$) which are mapped to gate G5. However, the degree of freedom of $G_5$ (OR gate with output 1) is only apparently 3, since the subcircuit which is rooted in $G_4$ describes a contradiction. Subsequently $y_4$ equals always 0. For an OR logic gate with output 1 ($y_5$) and input 0 ($y_4$) the degree of freedom is 1, and the only possible assignment is $x_1$=1 (row 17).

So, it is possible to realize a random assignment for one of the inputs mapped to a logic gate with freedom degree 2 or 3 *if and only if* the circuit does not contain subcircuits which describe tautologies or contradictions.

- a circuit with only AND gates and OR gates does not contain subcircuits which describe tautologies or contradictions.

- a circuit with AND, OR and NOT gates, shaped as a tree, does not contain subcircuits which describe tautologies or contradictions.

- a circuit with AND, OR and NOT gates, in which the NOT gates do not belong to any ramifications (see figure 6) does not contain subcircuits which describe tautologies or contradictions.

## 6. Recursive function for pre-processing

In order to guarantee that the *apparent* degree of freedom of each logic gate is also the *effective* one, the determination of all subcircuits which describe tautologies or contradictions is necessary to update correctly the values in data structures.

In order to do that it is possible to apply a simple recursive function to each gate, beginning from $G_m$ (with $0 \leq k \leq m$):

- Base case: if the inputs of gate $G_k$ are $x_i$, then we apply the heuristic to $G_k$ in order to determine if the elementary subcircuit, which is rooted in $G_k$, is a tautology or a contradiction itself;

- Recursion step: if one or both inputs of the gate $G_k$ are not $x_i$, then we call upon them the recursive function. As the calls comes back and if a value for $y_k$ has not been fixed yet, then we apply the heuristic so that we can determine if the circuit, which is rooted in $G_k$, is a tautology or a contradiction.

At this point let us explain the demonstration of correctness for the recursive function just introduced.

An elementary circuit (AND or OR gate, whose inputs are $x_i$) doesn't contain any subcircuit which describes a tautology or a contradiction (figure 7).

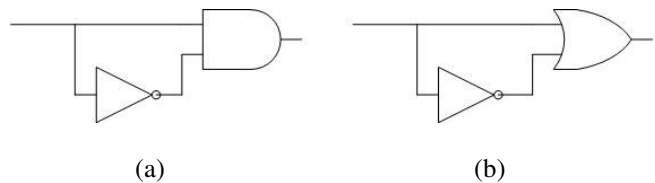

(a)                           (b)

**Figure 7.** Contradiction (a); Tautology (b).

In order to prove if a circuit is a tautology or a contradiction itself:

- Case "contradiction" (circuit rooted in an AND gate):

    Let the circuit output be 1 ($y_k=1$). Let us apply the heuristic. If we obtain a solution, then the circuit isn't a contradiction and therefore we simply come back from the call. On the other hand, if we obtain an inconsistent assignment we have determined a contradiction; in this case we should update the circuit truth table ($y_k=0$) and the degree table. At this point we propagate this information applying step 3.1 of the heuristic, if possible, and in the end we go back to the call.

- Case "tautology" (circuit rooted in an OR gate)

    Let the circuit output be 0 ($y_k=0$). Let us apply the heuristic. If we obtain a solution, then the circuit isn't a tautology and we go back to the call. However, if we get an inconsistent assignment then we have determined a tautology. In this case we should update the circuit truth table ($y_k=1$) and its degree table. We propagate this information implementing step 3.1 of the heuristic, if possible, and after that we go back to the call.

In so doing each subcircuit will be checked, being sure that it doesn't contain in its turn any subcircuits which describe tautologies or contradictions. In fact, if this should happen, the subcircuits should have already been detected thanks to the previous recursive calls.

Applying this function from $G_m$, it is possible to determine if the same circuit C describes a tautology or contradiction.

While applying the recursive function we proceed from a NOT gate straight to the AND or OR gate linked. If a contradiction or a tautology is rooted in an AND or OR gate automatically are fixed also the values of the NOT gate.

At the end of the pre-processing phase, if the circuit is neither a tautology nor a contradiction, we determine a solution applying the heuristic.

## 7. Final heuristic

Let us summarize all necessary steps in order to solve a general Circuit-SAT problem instance:

1. create data structures: circuit truth table and degree table;

2. pre-processing, apply recursive function (starting from $G_m$) in order to determine all sub-circuits which describe tautologies or contradictions;

3. if the circuit isn't a tautology or a contradiction:

   3.1. assign a value to the circuit output, $y_m=1$;

   3.2. delete from the circuit truth table all rows in which $y_m=0$; delete from degree table the values related to the index of previous deleted rows in circuit truth table;

   3.3. until unassigned $x_i$ exist:

      3.3.1. assign missing value to the inputs or output of all gates with degree of freedom 1; update circuit truth table and degree table (as shown in step 3.2);

      3.3.2. if there are no more gates with degree of freedom 1, assign the value 0 or 1 to an input of a gate with degree of freedom 2 or 3; then update the circuit truth table and degree table (as shown in step 3.2);

4. read values linked to the $x_i$;

The heuristic ends when all $x_i$ have been assigned: at each step (3.3.1 and 3.3.2) the heuristic assigns a value to an $y_j$, after $m$ assignments all $x_i$ have been assigned. The resulting assignment for the $x_i$ represents the sequence of 0 and 1 which satisfies the circuit C.

## 8. Computational complexity

The circuit truth table requires space *O(nm+m²)* and time *O(nm+m²)*. If we assume that, while creating the table initially all the values are set to null, it follows that a constant number of columns (that is 3) for each row is valued. This requires time *3·O(4m)=O(m)*. The degree table requires a time equal to *O(m)*.

Considering a circuit without subcircuits which describe tautologies or contradictions, it follows that the heuristic assigns to a $y_j$ a value at each step, that is it deletes some rows in the circuit truth table. The time required is *O(4m)=O(m)*.

The pre-processing phase calls the heuristic *m* times. Optimizing the recursive call, that is not applying the function to sub-circuits which are common to different calls, it follows that the time in the worst case is *m·O(m)=O(m²)*.

In case of a general Circuit-SAT instance, the necessary time in order to determine a solution is given by:

$$O(m) + O(m) + O(m^2) = O(m^2) \tag{4}$$

# References

[1] Cook, C.S. (1971) *The complexity of theorem-proving procedures*, 3rd ACM Symposium on Theory of Computing, ACM, New York, pp. 151-158

[2] Joshua J. Arulanandham, Cristian S. Calude, and Michael J. Dinneen. *Solving SAT with bilateral computing*. Report CDMTCS-199, Centre for Discrete Mathematics and Theoretical Computer Science, University of Auckland, Auckland, New Zealand, December 2002.